\documentclass[a4paper]{jpconf}

\usepackage{amssymb,amstext}

\newcommand{\bra}[1]{{\langle{#1}\vert}}
\newcommand{\ket}[1]{{\vert{#1}\rangle}}

\renewcommand{\Tr}{\mathop{\rm Tr}\nolimits}

\newcommand{\del}{\partial}

\newcommand{\cD}{{\cal D}}
\newcommand{\cH}{{\cal H}}
\newcommand{\cM}{{\cal M}}

\begin{document}

\title{Gravity-Matter Entanglement in Regge Quantum Gravity}

\author{Nikola Paunkovi\'c$^{1}$ and Marko Vojinovi\'c$^{2}$}

\address{$^1$SQIG --- Security and Quantum Information Group, Instituto de Telecomunica\~coes, and Departamento de Matem\'atica, Instituto Superior T\'ecnico, Universidade de Lisboa, 
Avenida Rovisco Pais 1049-001, Lisboa, Portugal}
\address{$^2$Grupo de F\'isica Matem\'atica, Faculdade de Ci\^ encias da Universidade de Lisboa, Campo Grande, Edif\'icio C6, 1749-016  Lisboa, Portugal}

\ead{npaunkov@math.ist.utl.pt, vmarko@ipb.ac.rs}

\begin{abstract}
We argue that Hartle--Hawking states in the Regge quantum gravity model generically contain non-trivial entanglement between gravity and matter fields. Generic impossibility to talk about ``matter in a point of space'' is in line with the idea of an emergent spacetime, and as such could be taken as a possible candidate for a criterion for a plausible theory of quantum gravity. Finally, this new entanglement could be seen as an additional ``effective interaction'', which could possibly bring corrections to the weak equivalence principle.
\end{abstract}

\textbf{Introduction.} 
The unsolved problems of interpreting quantum mechanics (QM) and formulating quantum theory of gravity (QG) are arguably the two most prominent ones of the twentieth century theoretical physics. Up to date, most of the efforts to solve the two were taken independently. Indeed, the majority of the interpretations of QM do not involve explicit dynamical effects (with notable exceptions of the spontaneous collapse and the de Broglie--Bohm theories), while the researchers from the QG community often adopt the many-world interpretation of QM. Nevertheless, the two problems share a number of similar unsolved questions and counter-intuitive features, such as nonlocality: entanglement-based 
quantum nonlocality, as well as the anticipated explicit dynamical nonlocality in QG (a consequence of quantum superpositions of different gravitational fields, i.e., different spacetimes and their respective causal orders). 
We analyse the generic entanglement between gravitational and matter fields in the Regge model of quantum gravity, and its possible impact to the fundamental questions regarding QM and QG.

\textbf{Regge quantum gravity model.} A simple toy model of quantum gravity with matter fields is the Regge quantum gravity with one real scalar field, whose construction can be motivated by the Loop Quantum Gravity research program \cite{jedan,dva}. The path integral of the model is 
\begin{equation} \label{ReggeStateSuma}
Z_{T} = \int \cD L \int \cD \Phi \;\; \exp \Big[ iS_{\rm Regge}(L) + iS_{\rm matter}(L,\Phi) \Big]\,,
\end{equation}
where $L$ are the lengths of the edges of the triangulation $T$ of a $4$-manifold $\cM_4$, and $\Phi$ are the values of the scalar field in $4$-simplices of $T$. The measure terms $\cD L$ and $\cD \Phi$ are defined via discretization induced by $T$. The actions $S_{\rm Regge}$ and $S_{\rm matter}$ represent lattice discretizations of the Einstein--Hilbert action for gravity and an action of the scalar field coupled to gravity, respectively. See~\cite{tri,cetiri} for details.

A generic state of the system, belonging to the kinematical Hilbert space $\cH_G\otimes \cH_M$, is 
\begin{equation} \label{TalasnaFunkcija}
\ket\Psi = \int\cD l \int\cD \phi \;\; \Psi( l, \phi) \;\; \ket{l} \ket{\phi}\,.
\end{equation}
However, since gravity is a theory with constraints, not every kinematical state is allowed, so we must choose the coefficients $\Psi(l,\phi)$ such that $\ket{\Psi}$ is an element of the physical Hilbert space $\cH_{\rm phys} \subset \cH_G\otimes \cH_M$. One such class of states are the Hartle--Hawking (HH) states \cite{pet}, defined by the following choice of the coefficients, for a given triangulation $T$:
\begin{equation} \label{HartleHawkingStanje}
\Psi(l,\phi) = \Psi_{\rm HH}(l,\phi) \equiv  \int\cD L \int \cD \Phi \;\; \exp \Big[ iS_{\rm Regge}(L,l) + iS_{\rm matter}(L,\Phi,l,\phi) \Big]\,.
\end{equation}
This expression differs from (\ref{ReggeStateSuma}) in that the triangulation $T$ is now assumed to have a nontrivial $3$-dimensional boundary $\del T$, and that the variables $l,\phi$ living on the boundary are not integrated over, in contrast to the bulk variables $L$ and $\Phi$.

Using (\ref{TalasnaFunkcija}) and (\ref{HartleHawkingStanje}), one can calculate the reduced density matrix of the Hartle--Hawking state,
$$
\hat \rho_M
= \Tr_G \ket\Psi  \bra\Psi
= \int\cD \phi \int \cD \phi' \; \; \left[ \int\cD l \;\; \Psi_{\rm HH}(l,\phi) \Psi^*_{\rm HH}(l,\phi') \right] \;\; \ket\phi \bra{\phi'}\,,
$$
where the integral in the brackets can be denoted as $Z_{T\cup\bar{T}}(\phi,\phi')$. The resulting density matrix can then be tested for entanglement by checking if the trace of its square equals one~\cite{sest}. In the Regge quantum gravity model the path integrals reduce to a finite number of ordinary integrals, which can then in principle be evaluated. For a generic triangulation, we obtain
$$
\Tr_M \hat\rho^2_M = \int\cD \phi \int\cD \phi' \;\; \left| Z_{T\cup\bar{T}}(\phi,\phi') \right|^2 \neq 1 \,,
$$
i.e., the gravitational and scalar degrees of freedom in the generic HH
state are entangled.

\textbf{Discussion.} In~\cite{sedam} Penrose argues that gravity-matter entanglement is at odds with (cla\-ss\-i\-cal) spacetime, seen as a (four-dimensional) differentiable manifold. In light of this, our result could be seen as a quantitative indicator that in quantum gravity one cannot talk of ``matter in a point of space'', i.e., this result could be seen as a confirmation of a ``spacetime as an emergent phenomenon''.
Thus, generic gravity-matter entanglement could be seen as a possible candidate for a criterion for a plausible theory of quantum gravity.
Entanglement is in standard quantum mechanics a generic consequence of the interaction. This new entanglement can be regarded as a consequence of an effective interaction (such as the ``exchange interactions'', which are a consequence of quantum statistics). This additional ``effective interaction'' can potentially lead to corrections to the weak equivalence principle.

\ack

NP was partially supported, under the CV-Quantum internal project at IT, by FCT PEst-OE/EEI/LA0008/2013 and UID/EEA/50008/2013. 
MV was supported by the FCT grant SFRH/BPD/46376/2008, the FCT project PEst-OE/MAT/UI0208/2013, and partially by the project ON171031 of the Ministry of Education, Science and Technological Development, Serbia.



\end{document}